\documentstyle[onecolumn,epsfig]{mn}

\def\be{\begin{equation}}
\def\ee{\end{equation}}
\def\lme{\log M_e}
\def\lre{\log R_e}
\def\lmf{\overline{\log M}_f}

\def\lm{\overline{\log M}}
\def\mfmi{M_{GCS,f}/M_{GCS,i}}
\def\nfni{N_{f}/N_{i}}
\def\dmio{\Delta \overline{\log M}_{inn-out}}
\def\sgf{\sigma_f}

\def\ltorder{\hbox{ \rlap{\raise 0.425ex\hbox{$<$}}\lower
0.65ex\hbox{$\sim$} }} 
\def\gtorder{\hbox{ \rlap{\raise 0.425ex\hbox{$>$}}\lower
0.65ex\hbox{$\sim$} }}

\begin{document}
\title{Evolution of globular cluster systems in elliptical
galaxies. II: power-law initial mass function}

\author[E.Vesperini]{E.Vesperini\thanks{E-mail:vesperin@falcon.astro.umass.edu
}
\\ Department 
of Astronomy, University of Massachusetts, Amherst, MA, 01003, USA}
\maketitle
\begin{abstract}
We have studied the evolution of globular cluster systems (GCS) in elliptical
galaxies with a power-law initial GCS mass function (GCMF) ($f(M)
\propto M^{-\alpha}$)
similar to that predicted by some theoretical studies of globular
clusters formation and to that of young cluster systems observed in
merging galaxies. 

We have carried out a survey over a large number of different host
galaxies and we have considered
different values for the index, $\alpha$, of the initial power-law
GCMF ($\alpha=1.5,1.8,2.0$); we show the dependence of the main GCS
final properties 
(mean mass and dispersion of the final GCMF, fraction of surviving
clusters, radial gradient of the GCMF parameters) on the
structure of the host galaxy and on the slope of the initial GCMF. 

For a subsample of host galaxies with values of effective 
masses and radii equal to those determined using observational data for a
number of giant, normal and dwarf galaxies our results show
that the relation between the final GCMF
properties and those of the host galaxies as well as the dependence of
the final GCMF parameters on the galactocentric distance within
individual galaxies differ from those observed in old GCS: the values
of the final GCS mean mass are in general smaller ($4.2\ltorder
\lmf\ltorder 5.0$) than those observed and the galaxy-to-galaxy
dispersion of $\lmf$ is larger than that reported by observational analyses. 
The results are compared with those of a companion paper in which we
investigated the evolution of GCS with a log-normal initial GCMF and
in which the final GCS properties were perfectly consistent with
observations. 
\end{abstract}
\section{Introduction}
In a recent paper (Vesperini 2000) we have investigated the
evolution of the main 
properties of globular clusters systems (hereafter GCS) in elliptical galaxies 
starting with a log-normal GCS initial mass
function. The choice of such a functional form for the initial mass
function was motivated  by the fact
that all the globular cluster system mass functions (hereafter
GCMF; we will indicate a globular cluster system luminosity function
by GCLF) observed so far are well fitted by a log-normal distribution; in
particular, a log-normal GCMF fits well the
observed GCMF of clusters located in the external regions of galaxies
where evolutionary processes are unlikely to have significantly
altered the initial properties of clusters.

In Vesperini (2000) it was shown that the main final GCS properties
resulting from the evolution of GCS with a log-normal initial GCMF and the
relations 
between GCS properties and those of the host galaxies are in very
good agreement with those reported by a number of observational analyses. 

In this paper we study the evolution of GCS starting with a power-law
initial GCMF. The motivation  for considering this functional form for
the initial GCMF comes
both from  some theoretical investigations of
globular clusters formation (see e.g. Elmegreen \& Efremov 1997, Harris
\& Pudritz 1994) which predict this  shape  for 
 the initial GCMF and from several observational
studies of young cluster systems in interacting and merging
galaxies (see e.g. Schweizer et al. 1996, Miller et al. 1997, Johnson
et al. 1999, Zepf et al. 1999; see also Whitmore 1999 for a review and
references therein) showing that these systems are characterized by a
power-law luminosity function.  
In fact, whether the observed power-law luminosity function
indeed corresponds to an underlying power-law mass function or it results
from the age spread of clusters with a log-normal mass function is
matter of debate (Fritze-von Alvensleben 1998, 1999, Carlson et
al. 1999, Zhang \& Fall 1999); moreover a recent analysis of the GCLF
of young clusters in the Antennae system seems to show a turnover in
the LF which would be therefore better fitted by a two-index power-law
(Whitmore et al. 1999).  

Most observational analyses, on the basis of  the high luminosities and
compact sizes observed, 
support the idea that the young clusters found in interacting
galaxies are globular clusters
which will eventually evolve into  systems 
similar to old globular clusters 
(see e.g. the results of  Ho \& Filippenko 1996a, 1996b which strongly
suggest that at least the brightest clusters must indeed be globular
clusters; see also Carlson et al. 1998, 1999)
 but  this point has been
questioned too: van den Bergh (1995), on the basis of the shape of the
GCLF, has claimed that these objects would be more properly classified
as young open clusters rather than as young globular clusters.

Several theoretical investigations have shown that evolutionary
processes can lead to the disruption of a significant number of
clusters and alter the parameters and shape of the initial GCMF (Fall
\& Rees 1977, Fall \& Malkan 1978, Caputo \& Castellani 1984,
Chernoff, Kochanek \& Shapiro 1986, Chernoff \& Shapiro 1987, Aguilar,
Hut \& Ostriker 1988, Vesperini,
1994, 1997, 1998, Okazaki \& Tosa 1995, Capuzzo Dolcetta \& Tesseri
1997, Gnedin \& Ostriker 1997, Murali \& Weinberg 1997a,
1997b, Ostriker \& Gnedin 1997, Baumgardt 1998); in particular a
power-law initial GCMF can be turned by evolutionary processes into a
log-normal GCMF  (or in general into a bell shaped GCMF). On the other hand
it is not clear whether evolutionary processes, acting with different 
efficiency in galaxies with different structures, can turn an initial
power-law into a log-normal GCMF with approximately universal
parameters and with a weak radial variation of the GCMF parameters
within individual galaxies as found in most current observational studies.
\begin{figure}
\centerline{\psfig{file=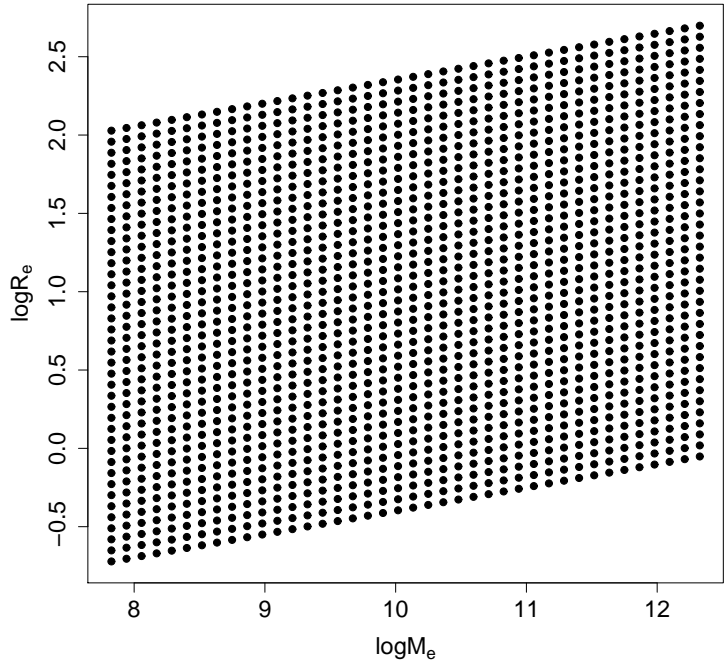,height=10cm,width=10cm,angle=0}}
\caption{Set of values of the effective mass, $M_e~(M_{\odot})$, and of the
effective radius, $R_e$ (kpc), of the host galaxies considered in the paper.}
\end{figure}

In Vesperini (2000) it has been shown
that, for a log-normal initial GCMF, the galaxy-to-galaxy variation of
the GCMF parameters resulting from theoretical calculations  is
perfectly consistent with the 
observed trends and that a considerable 
disruption of clusters does not necessarily give rise to a radial
gradient of the mean mass of clusters inconsistent with 
observations.
The goal of this paper is that of determining if the same conclusions
hold for the case of GCS with a power-law initial GCMF.

The layout of the paper is the following. In \S 2 we briefly sketch the method
used for the investigation and the initial conditions adopted; in \S 3 we
describe the general results while in \S 4 we discuss in detail the
implications of our results for galaxies for which observational data
are available. We summarize our conclusions in \S 5.

\section{Method and initial conditions}
The method adopted to follow the evolution of the masses of individual globular
clusters is the same used in Vesperini (2000; see also Vesperini 1998) and we
refer to that paper for further details. The evolutionary processes
included are mass loss due to stellar evolution, two-body relaxation,
the presence of the tidal field of the host galaxy and dynamical
friction; the effects of the time-variation of the tidal field for
clusters on non-circular orbits (see e.g. Weinberg 1994a, 1994b,
1994c, Gnedin, Hernquist \& 
Ostriker 1999) were not included in the N-body
simulations by Vesperini \& Heggie (1997) and are not considered here.
The  analytical expression used to determine the mass of a cluster 
located at distance $R_g$ from the center of its host galaxy at time $t$
is based on the N-body simulations carried out by Vesperini \& Heggie
(1997) and is given by  

\be
{M(t)\over M_i}=1-{\Delta M_{st.ev.}\over M_i}-{0.828\over F_{CW}}t.
\ee
 $t$ is time measured in Myr, ${\Delta M_{st.ev.}\over M_i}$ is the mass
loss due to stellar evolution (see eq.10 in Vesperini \& Heggie 1997);
$F_{CW}$ is a parameter introduced by Chernoff \& Weinberg (1990), which is
proportional to the initial relaxation time of the cluster and  is defined as 
\be
F_{CW}={M_i \over M_{\odot}}{R_g \over \hbox{kpc}}{1\over \ln N}{220
\hbox{km s}^{-1} \over v_c},
\ee
where $N$ is the total initial number of stars in the cluster and $v_c$ is the
circular velocity around the host galaxy. 
For the host galaxy we will adopt a simple isothermal model with constant
circular velocity.
The effects of dynamical friction at any time $t$ are included by removing, at
that time, all clusters with time-scales of orbital decay (see e.g. Binney \&
Tremaine 1987) smaller than $t$.

The values of the effective masses, $M_e$, and effective radii, $R_e$, of
the host galaxies considered are the 
same studied in Vesperini (2000) and they are plotted in Fig. 1.
\begin{figure}
\centerline{\psfig{file=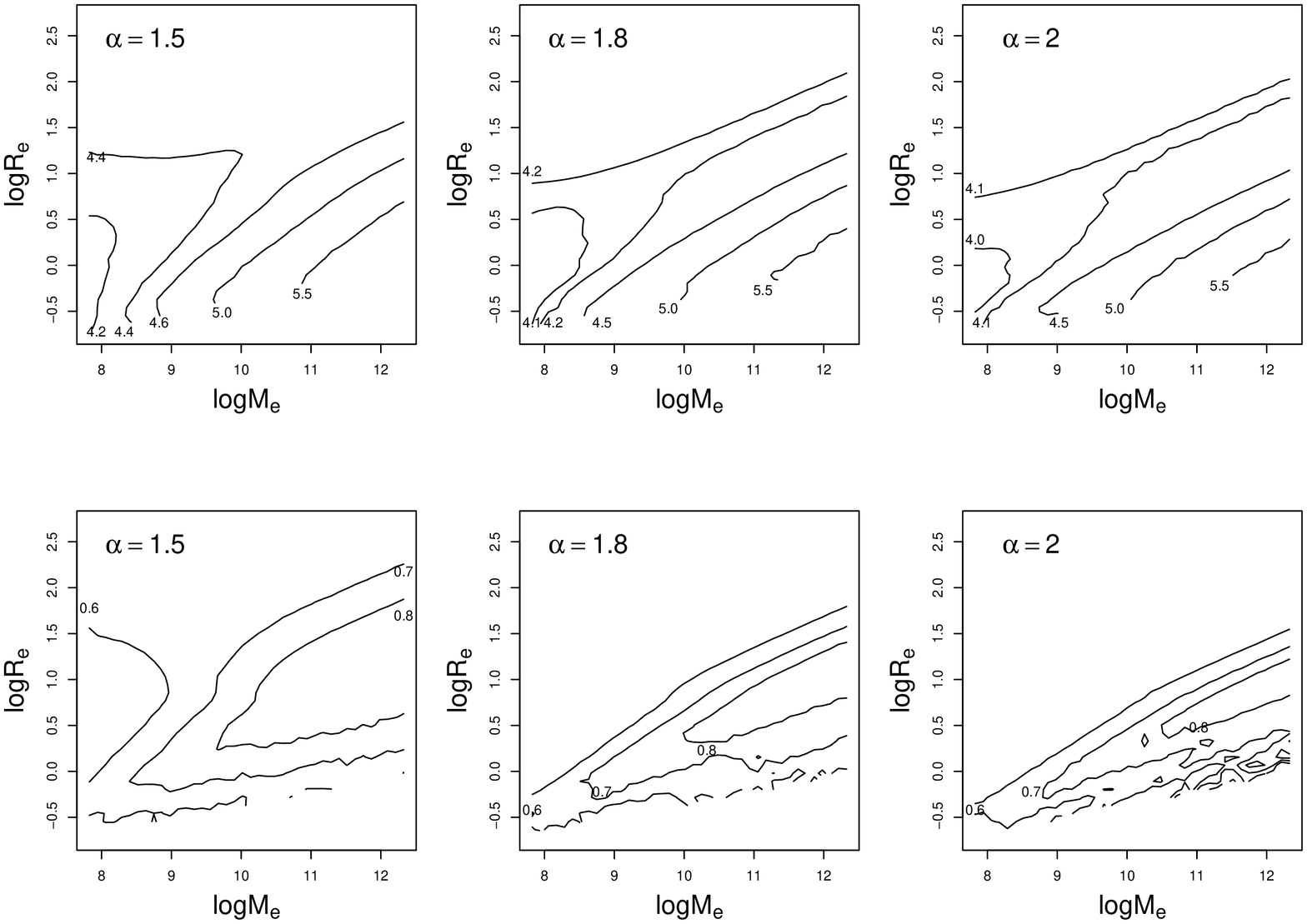,height=12cm,width=10cm,angle=-90}}
\caption{Contour plots of $\lmf$ (upper panels) and of $\sgf$ (lower
panels) in the plane $\lme-\lre$. The index  of the initial power-law 
GCMF, $\alpha$,  is shown on the upper left
of each panel.}
\centerline{\psfig{file=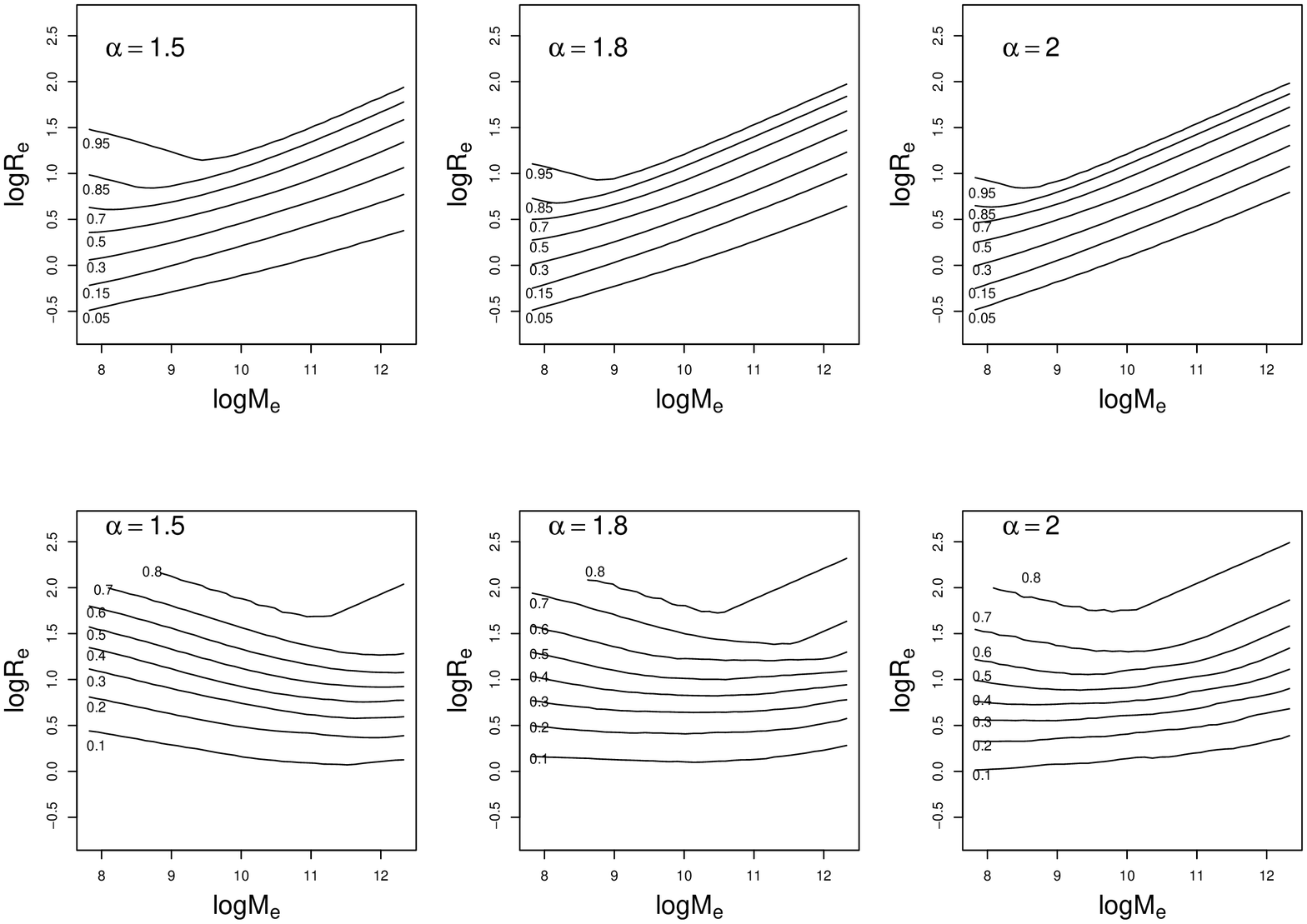,height=12cm,width=10cm,angle=-90}}
\caption{Contour plots of $\nfni$ (upper panels) and of $\mfmi$ (lower
panels) in the plane $\lme-\lre$. The index  of the initial power-law 
GCMF, $\alpha$,  is shown on the upper left
of each panel.}
\end{figure}

Each GCS investigated initially contains 20000 clusters with a GCMF given by 
\be
dN(M)=AM^{-\alpha} dM~ \hbox{for}~ 10^4~M_{\odot}<M<10^7~M_{\odot};
\ee
the values of $\alpha$ considered are  $\alpha=1.5,~1.8,~2.0$ which
are close to those determined from observation of the GCLF of young
cluster systems in merging and interacting galaxies.
Clusters are initially distributed inside the host galaxy between
$R_g=0.16R_e$ and $R_g=5R_e$ with a 
number of clusters per cubic kpc  proportional to $R_g^{-3.5}$ which
is similar to that observed for Galactic halo clusters (see also
Murali \& Weinberg 1997a where a similar slope for the initial radial
distribution is derived for the M87 GCS). 

The evolution of each GCS is followed for 15 Gyr.
 
\section{General results}
Fig. 2 shows the contour plots of the final mean mass and dispersion
of the GCMF, $\lmf$ and $\sgf$, in the plane
$\log M_e-\log R_e$
for the three values of $\alpha$ considered. The flatter the
initial GCMF, the larger $\lmf$ is
for a given model of the host galaxy.  We will discuss below in
section 4 the implications of 
our results for host galaxies with values of $M_e$ and $R_e$ equal to
those determined by observations but we note here that the range of 
values spanned by $\lmf$ is significantly larger than that obtained in
Vesperini (2000) for a log-normal 
initial GCMF (compare Fig.2a of Vesperini 2000 with the upper panels
of Fig. 2). 

Fig. 3 shows the contour plots of the fraction of the total initial
number of clusters surviving after 15 Gyr, $N_f/N_i$, and of the ratio
of the total mass of the survived clusters to the total initial mass
of all clusters, $M_{GCS,f}/M_{GCS,i}$, in the plane
$\log M_e-\log R_e$.
For host galaxies where evaporation by internal relaxation is 
the dominant disruption process, the fraction of surviving clusters is smaller
for steeper initial GCMF (in which the fraction of low-mass
clusters is larger); on the other hand, in galaxies where dynamical
friction is more 
important, the fraction of surviving clusters decreases  for initial GCMF with
smaller values of $\alpha$ which initially contain a larger  fraction of
high-mass clusters.   

The left panel of Fig. 4 shows  the contour plot of the difference
between the final mean mass of inner clusters ($R_g<R_e$) and outer
clusters ($R_g>R_e$) for $\alpha=1.8$ (hereafter we  
will focus our attention on this value of $\alpha$ which is equal  to
that found in several young GCS in merging and interacting galaxies): 
for a large number of different host galaxies evolutionary processes
lead to a significant radial gradient of the mean mass.

In general, GCS with initial GCMF containing a large fraction of
low-mass clusters, such as the power-law considered here or a log-normal
GCMF with a small initial value of $\lm$ (see Vesperini 1998), are
more prone to the formation of 
stronger radial gradients of $\lm$ which are inconsistent with the very
weak (if any) radial trends reported by observational analyses.
\begin{figure}
\centerline{\psfig{file=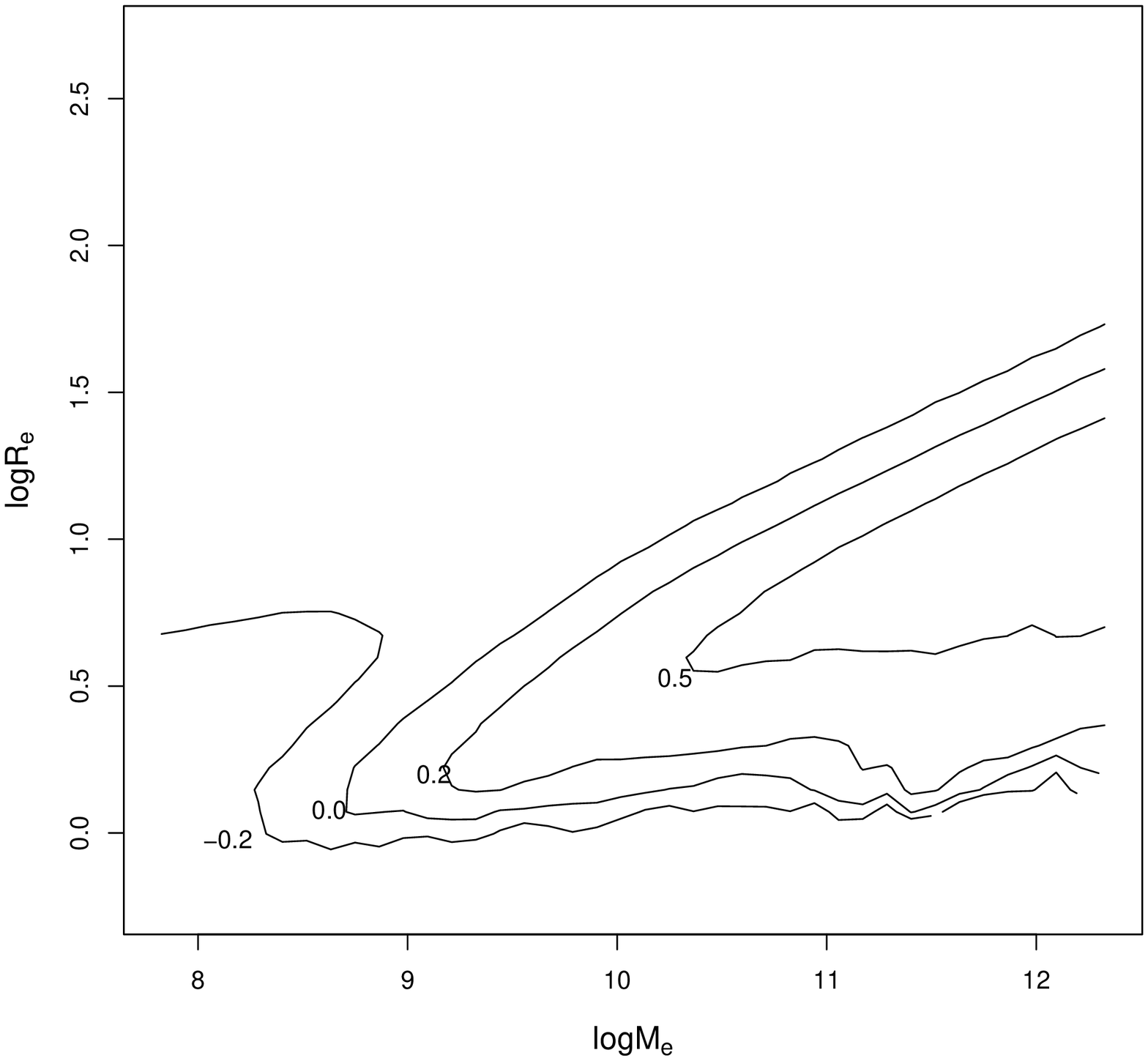,height=9cm,width=9cm,angle=0}
\psfig{file=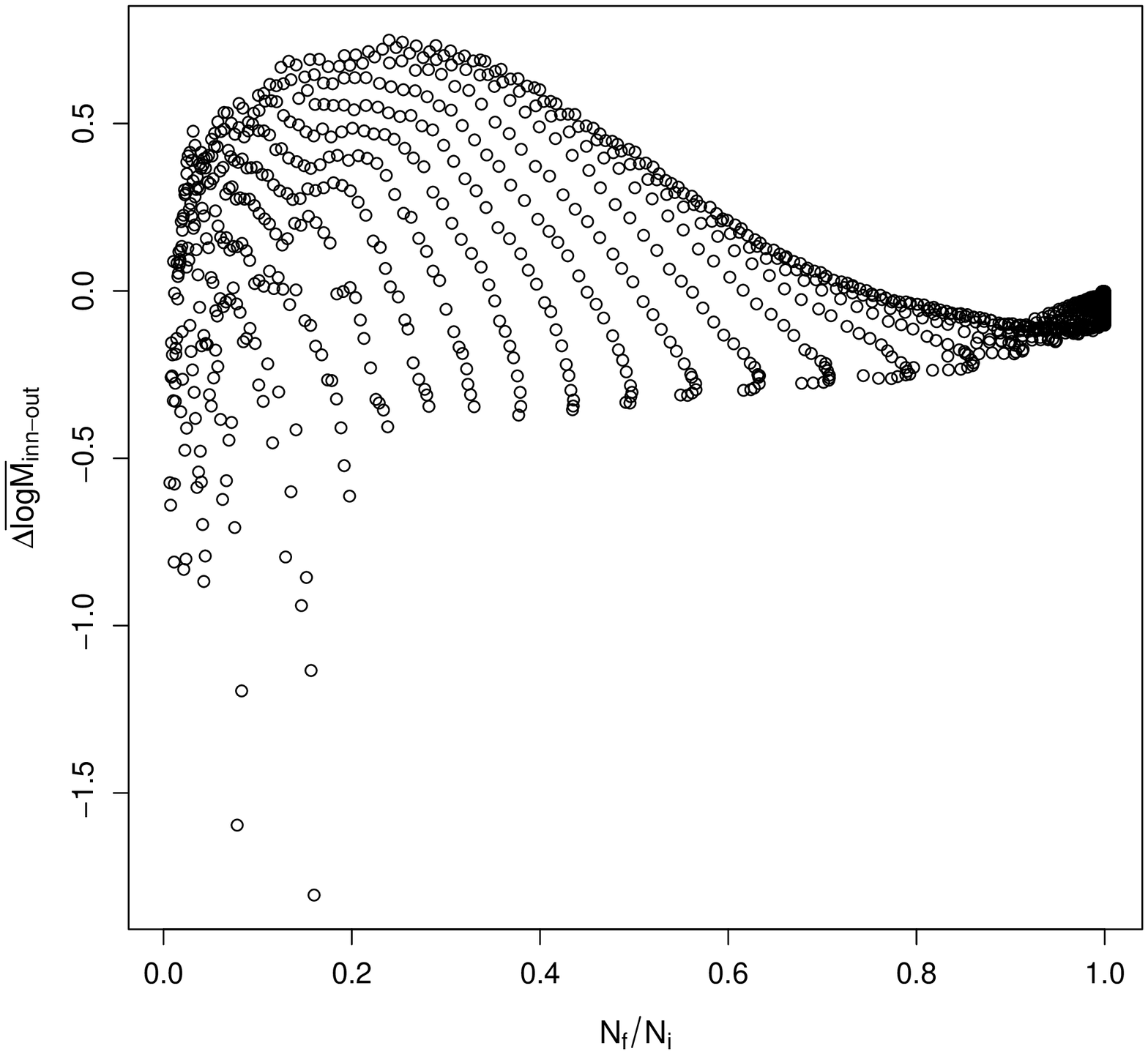,height=9cm,width=9cm,angle=0}}
\caption{(Left panel) Contour plot of the difference between  the final mean
 mass of inner ($R_g<R_e$)  and outer ($R_g>R_e$)  clusters, $\dmio$, in the plane 
$\lme-\lre$. (Right panel) $\dmio$ versus
the fraction of surviving clusters after 15 Gyr, $\nfni$. Both panels
refer to systems with an initial power-law GCMF with $\alpha=1.8$. }
\end{figure}
The right panel of Fig. 4 shows that, as it happens for a log-normal
initial GCMF (see 
Fig.4 in Vesperini 2000), a small
value of $\dmio$ does not necessarily imply
a negligible disruption of clusters; 
the difference in the range of values of  $\dmio$
between GCS with a power-law initial GCMF considered here and those
with an initial log-normal GCMF studied in Vesperini (2000) is not due to
a difference in the fraction of disrupted clusters (though, in
general, GCS with a power-law GCMF tend to have a smaller fraction of
surviving clusters) but rather to the different mass
distribution of disrupted clusters. 
\section{Implications}
In this section we
discuss the implications of our results for a sample of galaxies
for which effective masses and radii have been determined using
observational data by Burstein et al. (1997) (we adopted $H_0=75~ \hbox{km}~
\hbox{s}^{-1}~ \hbox{Mpc}^{-1}$).  
We will focus our attention on the results obtained for $\alpha=1.8$ but
the general trends found  are common to the other values of $\alpha$
we have studied. 
\subsection{Mean mass of clusters and fraction of surviving clusters}
Fig. 5 shows the contour plot of $\lmf$ already shown in Fig. 2
and discussed in \S 3 with the points corresponding to the observational values
of $\log M_e$ and $\log R_e$ superimposed. 
\begin{figure}
\centerline{\psfig{file=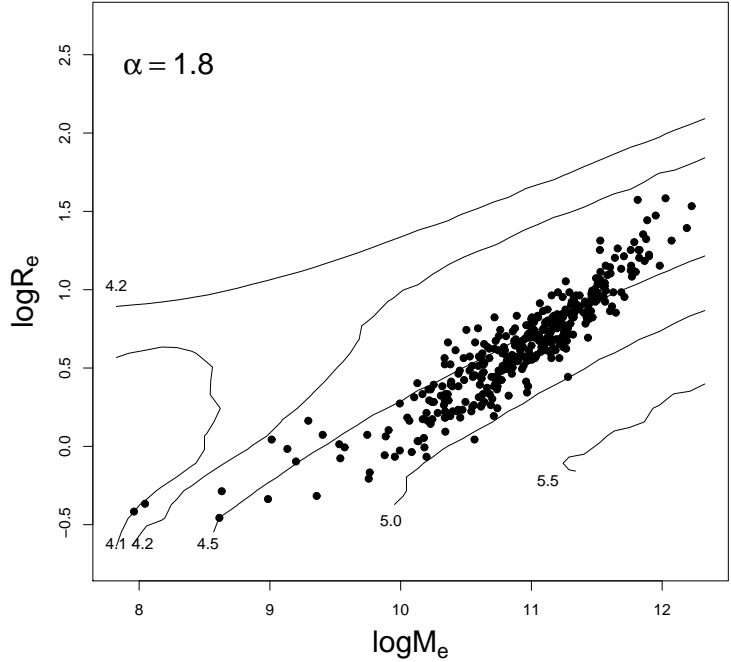,height=10cm,width=10cm,angle=0}}
\caption{Contour plot of $\lmf$ in the plane $\log M_e-\log R_e$
(already shown in the central upper panel of Figure 2) with 
observational values of $\log M_e$ and $\log R_e$ for elliptical galaxies 
(data from Burstein et al. 1997) superimposed as filled dots. The
initial GCMF adopted is a power-law function with
$\alpha=1.8$. }
\end{figure}
Figs 6a-c show $\lmf$, $N_f/N_i$, $\mfmi$ versus the observational
values of $\lme$.  

A recent observational analysis by Harris (2000) reports, for giant
ellipticals, a mean turnover magnitude $M_V^0=-7.33$ with a
galaxy-to-galaxy rms dispersion of 0.15 (corresponding, for $M/L_V=2$, 
to $\lmf\simeq 5.16$; the galaxy-to-galaxy rms dispersion of $\lmf$ is 0.06).
\begin{figure}
\centerline{\psfig{file=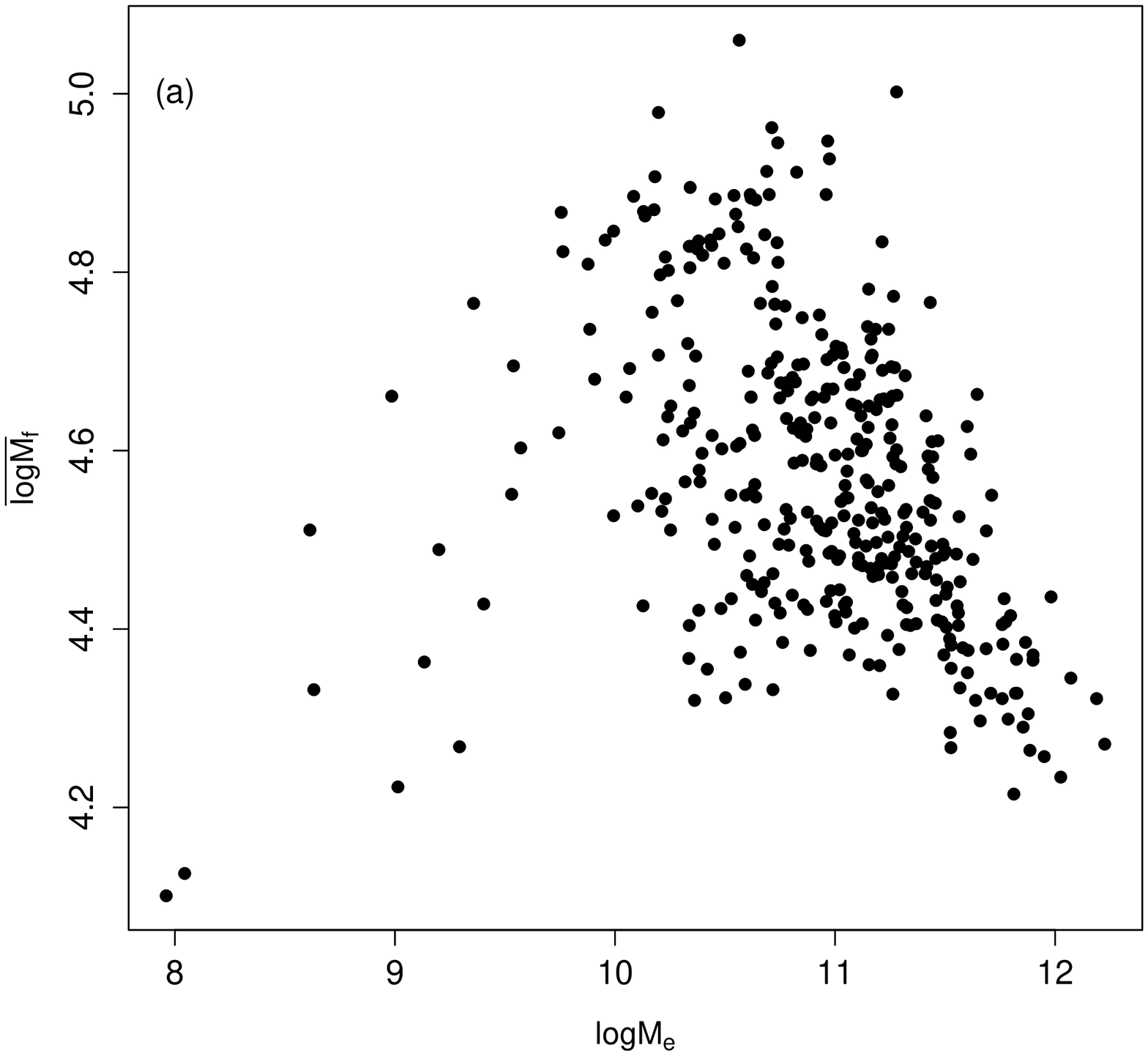,height=7cm,width=7cm,angle=0}}
\centerline{\psfig{file=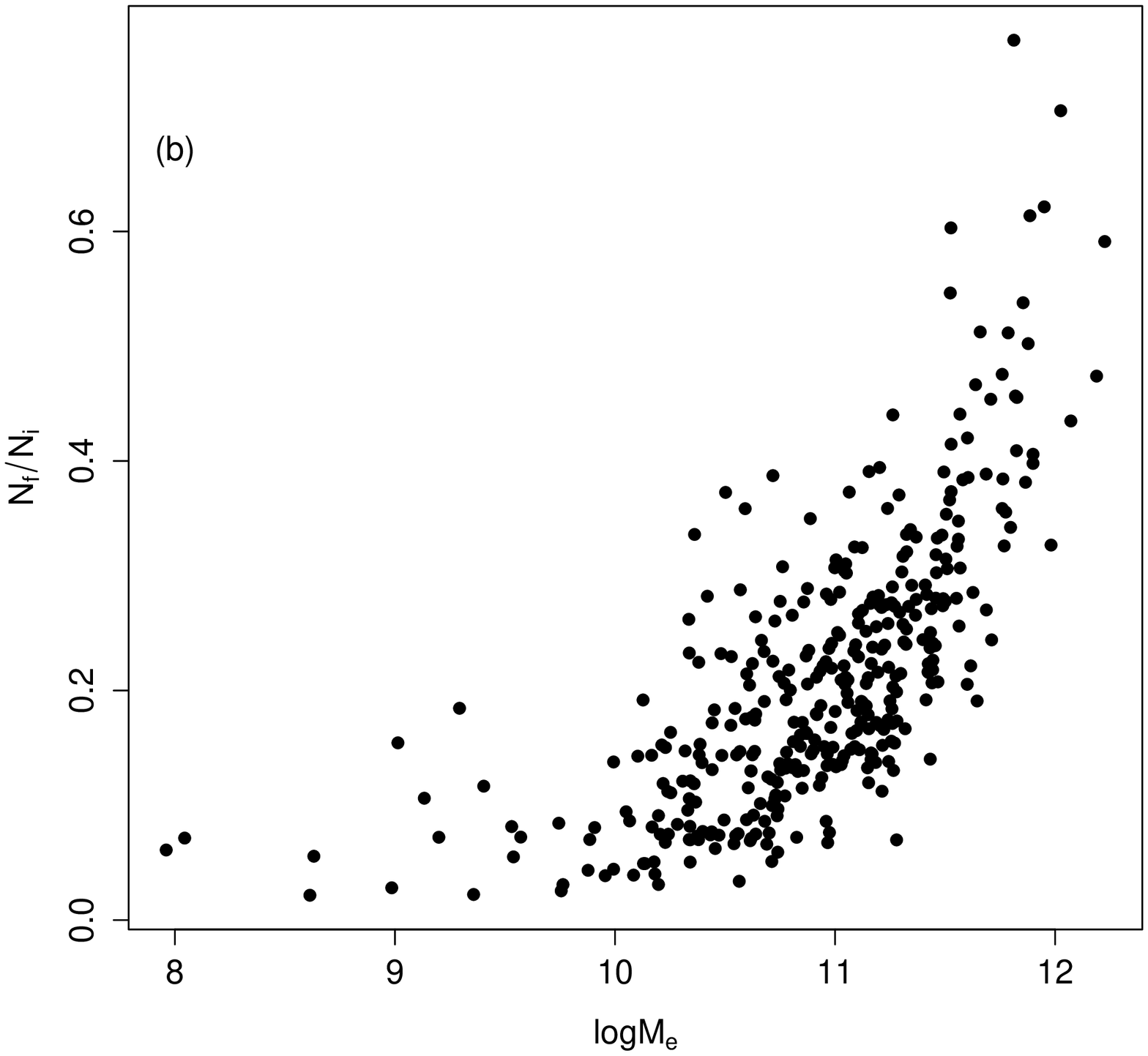,height=7cm,width=7cm,angle=0}}
\centerline{\psfig{file=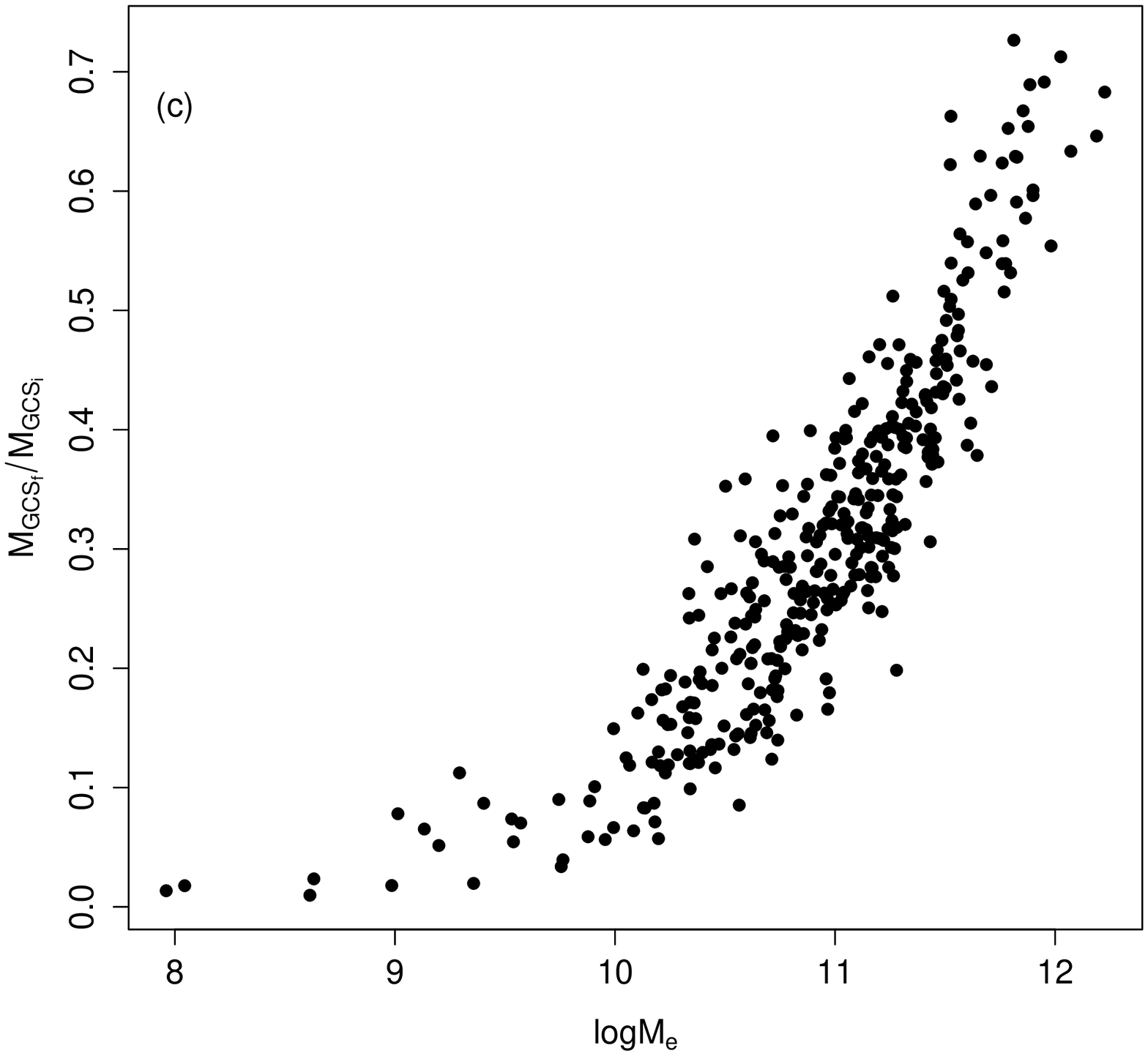,height=7cm,width=7cm,angle=0}}
\caption{(a) $\lmf$, (b) $\nfni$ and (c) $\mfmi$ versus the logarithm 
of the effective  
mass of  the host galaxy for a set of globular cluster systems located in host
galaxies with values of $R_e$ and $M_e$ equal to the observational values 
plotted in Fig. 5. The initial GCMF adopted is a power-law function
with $\alpha=1.8$. }
\end{figure}
For  $10\ltorder \log M_e\ltorder 12$, 
 Figs 5 and 6a show instead  
a large spread of values of $\lmf$ ($4.2<\lmf<5.0$) and a
general trend for  $\lmf$ to increase as $M_e$ decreases.
\begin{figure}
\centerline{\psfig{file=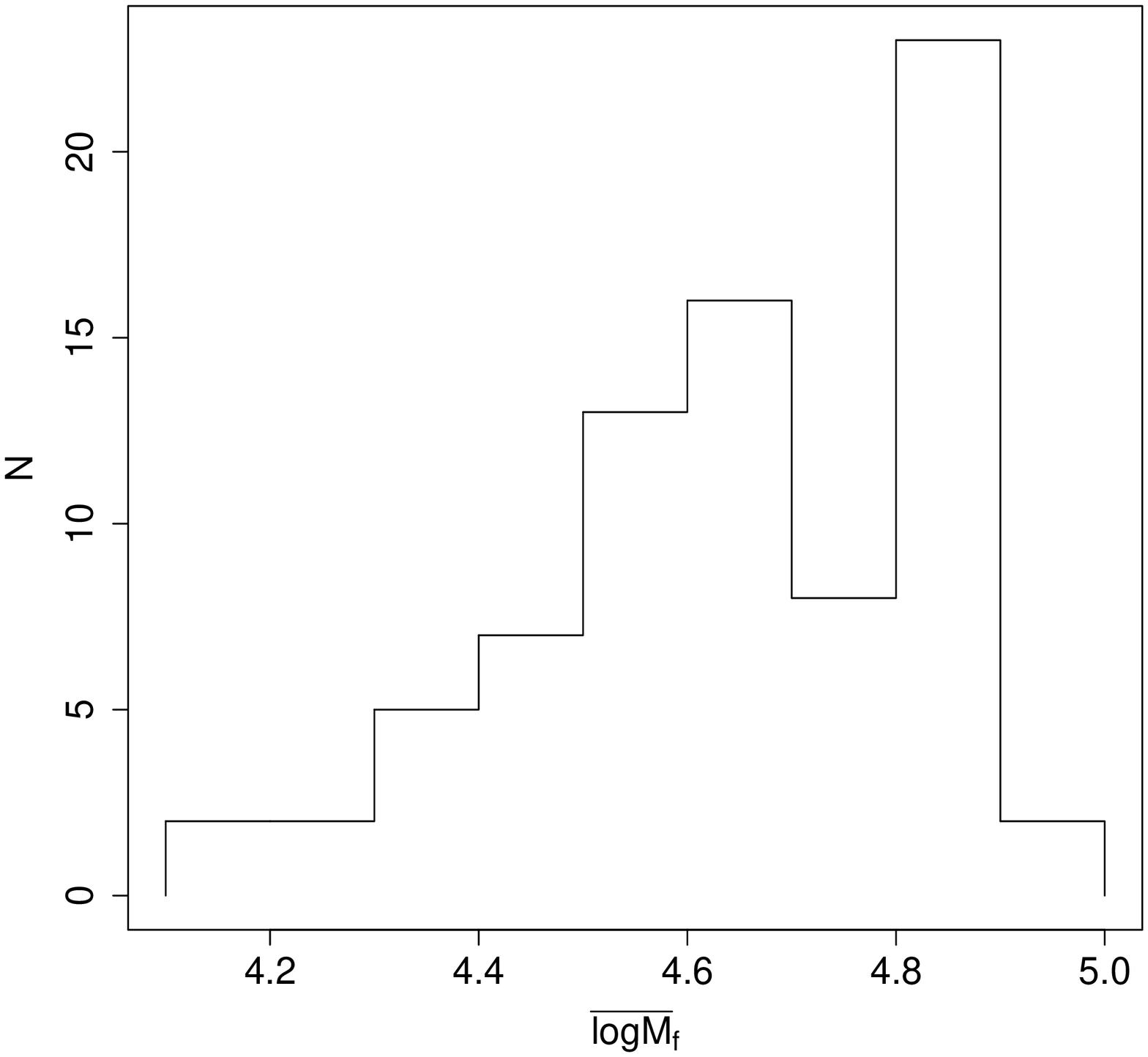,height=10cm,width=10cm,angle=0}}
\caption{Distribution of $\lmf$ from the simulations discussed in
section 4  for 
GCS located in host galaxies with values of $\log
R_e$ and $\log M_e$ equal to the observed values (data from Burstein et
al. 1997) for galaxies with $\lme>10.5$. 
The initial power-law GCMF adopted has $\alpha=1.8$.  }
\end{figure}

The inconsistency between the observational result of Harris (2000) and the
final properties of GCS starting with an initial power-law with
$\alpha=1.8$ is further illustrated by Fig.7: this figure shows the
distribution of theoretical values of $\lmf$ for 
globular cluster systems located in host galaxies with values of $\log
R_e$ and $\log M_e$ equal to the observational values reported by Burstein et
al. (1997) for galaxies with $\lme>10.5$. 

The four panels of Fig. 9 show the distribution of $\lmf$ for
low- ($\lme<9.5$), intermediate- ($9.5<\lme<10.5$) and
high-mass ($\lme>10.5$) ellipticals at $t=2,~5,~10,~15$ Gyr;
the values of $\log R_e$ and $\log M_e$ considered for these plots are
those shown in 
Fig. 8. The region of the $\lme-\lre$ plane shown in Fig. 8 is that
where most observational data fall.
\begin{figure}
\centerline{\psfig{file=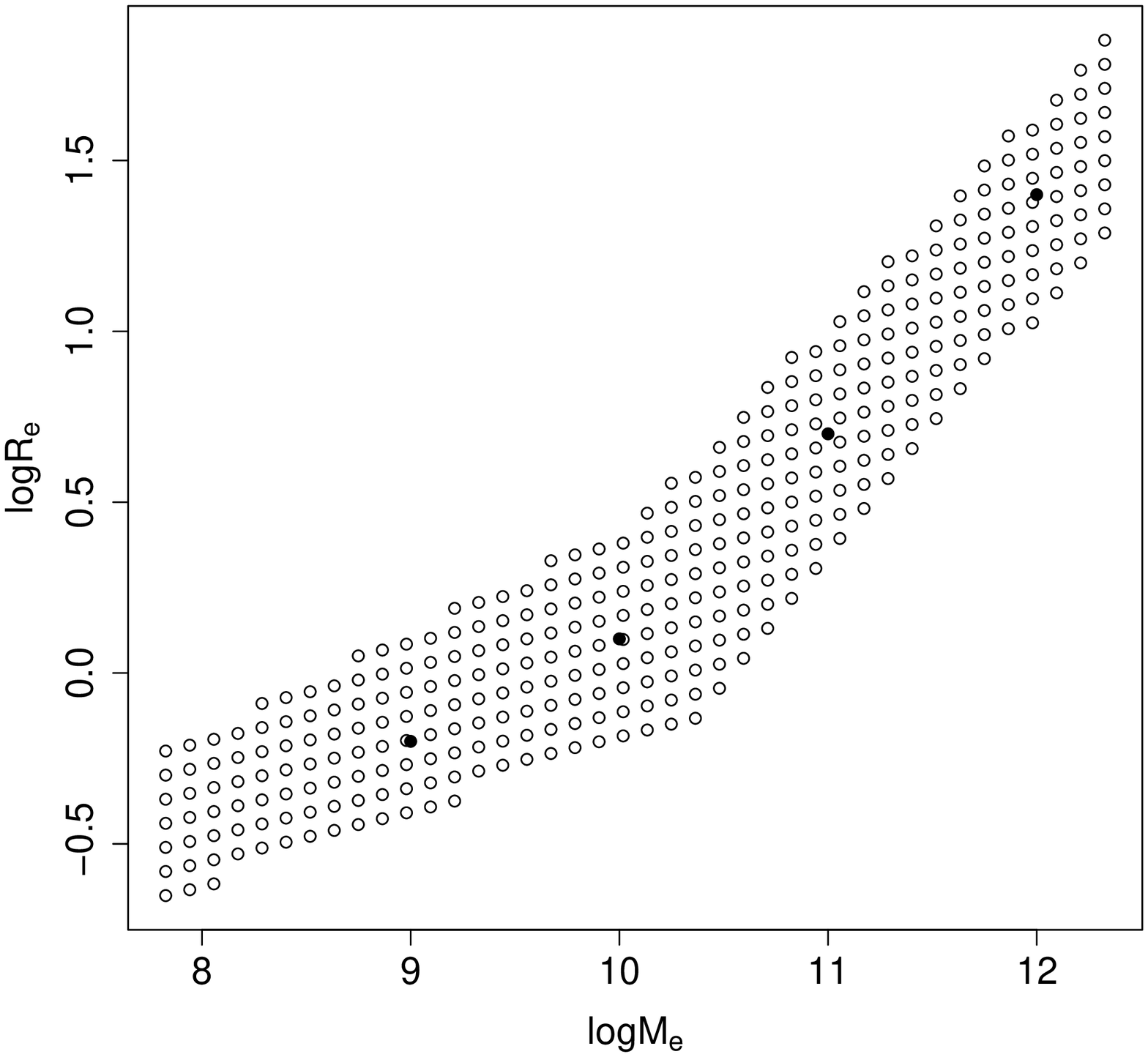,height=10cm,width=10cm,angle=0}}
\caption{ Values of $R_e$ and $M_e$ of host galaxies considered for
some of the simulations discussed in section 4.1 (see Fig.9).
Filled dots indicate the values of $\log M_e$
and $\log R_e$ for the host galaxies considered in \S 4.2 for a 
detailed study of the dependence of the main GCS properties on the
galactocentric distance. }
\end{figure}
\begin{figure*}
\centerline{\psfig{file=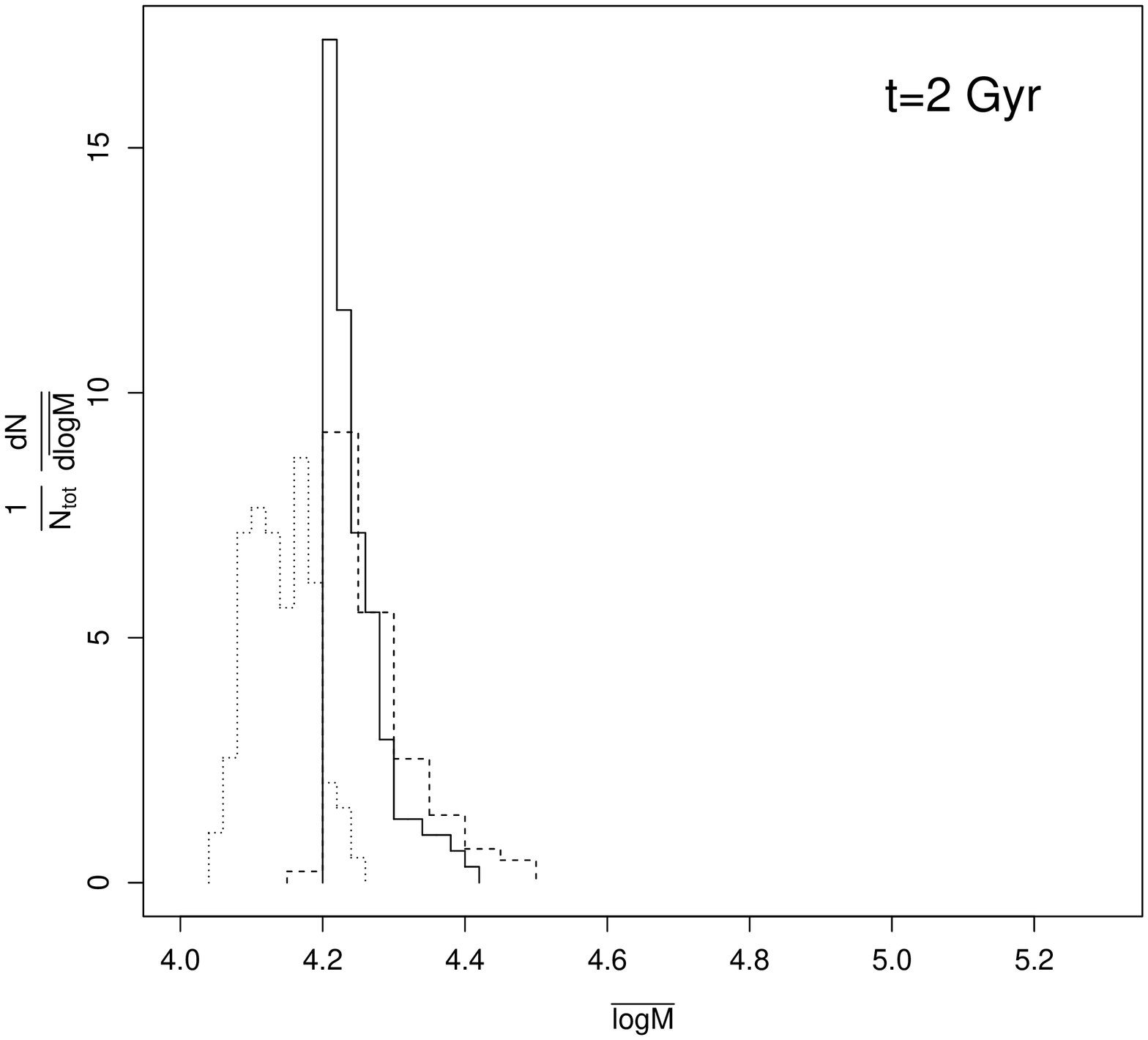,height=6cm,width=6cm,angle=0}
\psfig{file=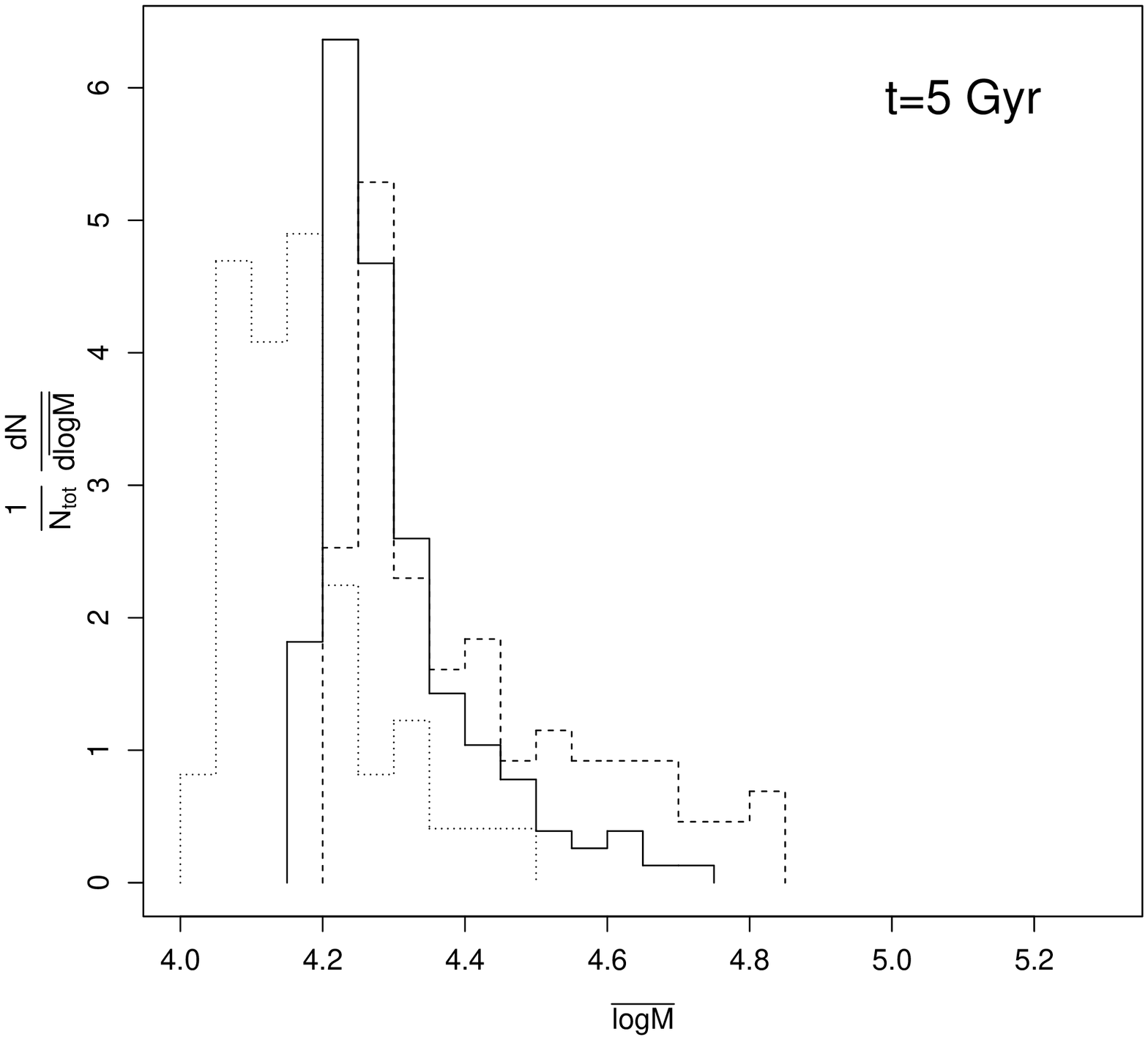,height=6cm,width=6cm,angle=0}}
\centerline{\psfig{file=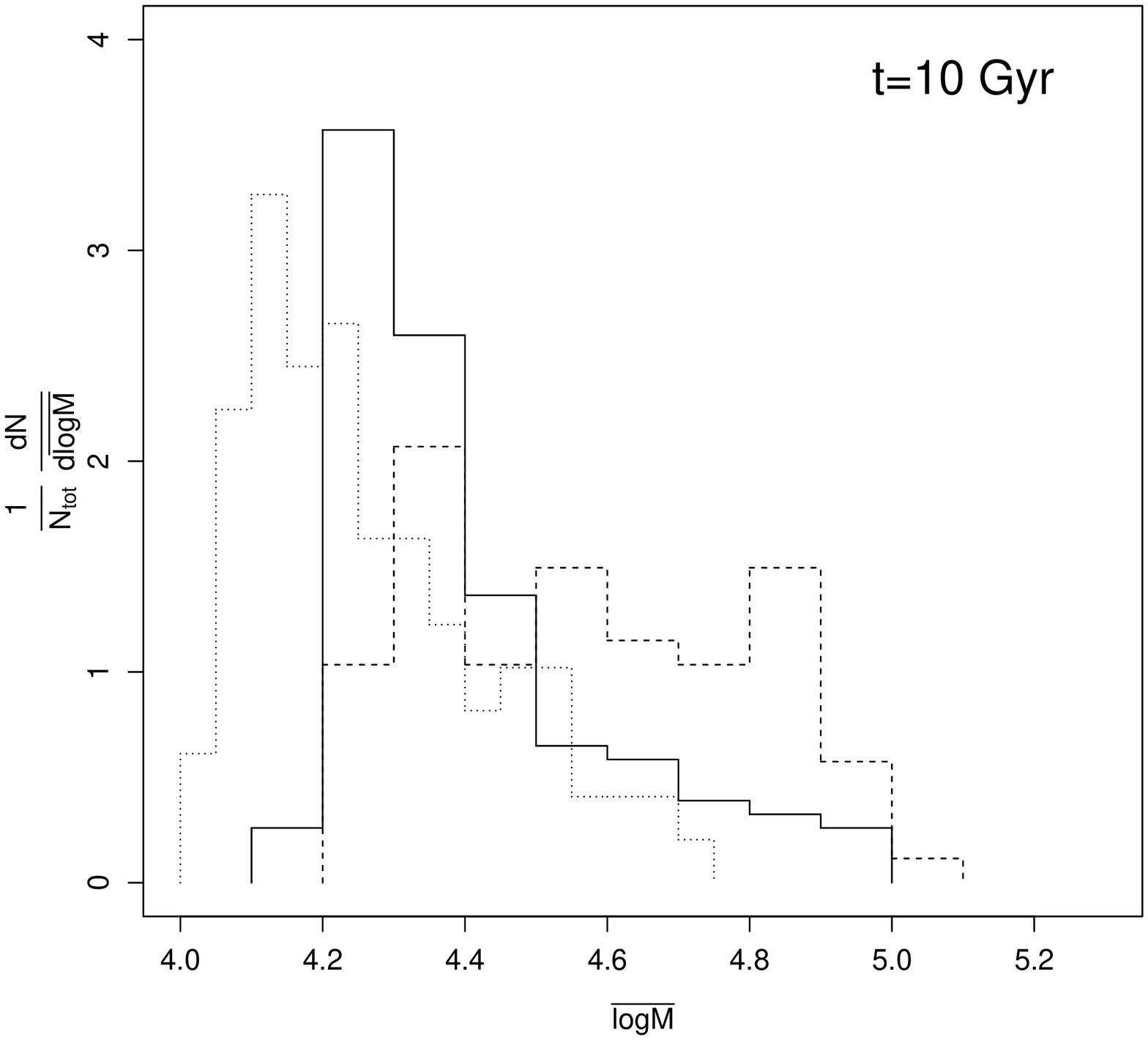,height=6cm,width=6cm,angle=0}
\psfig{file=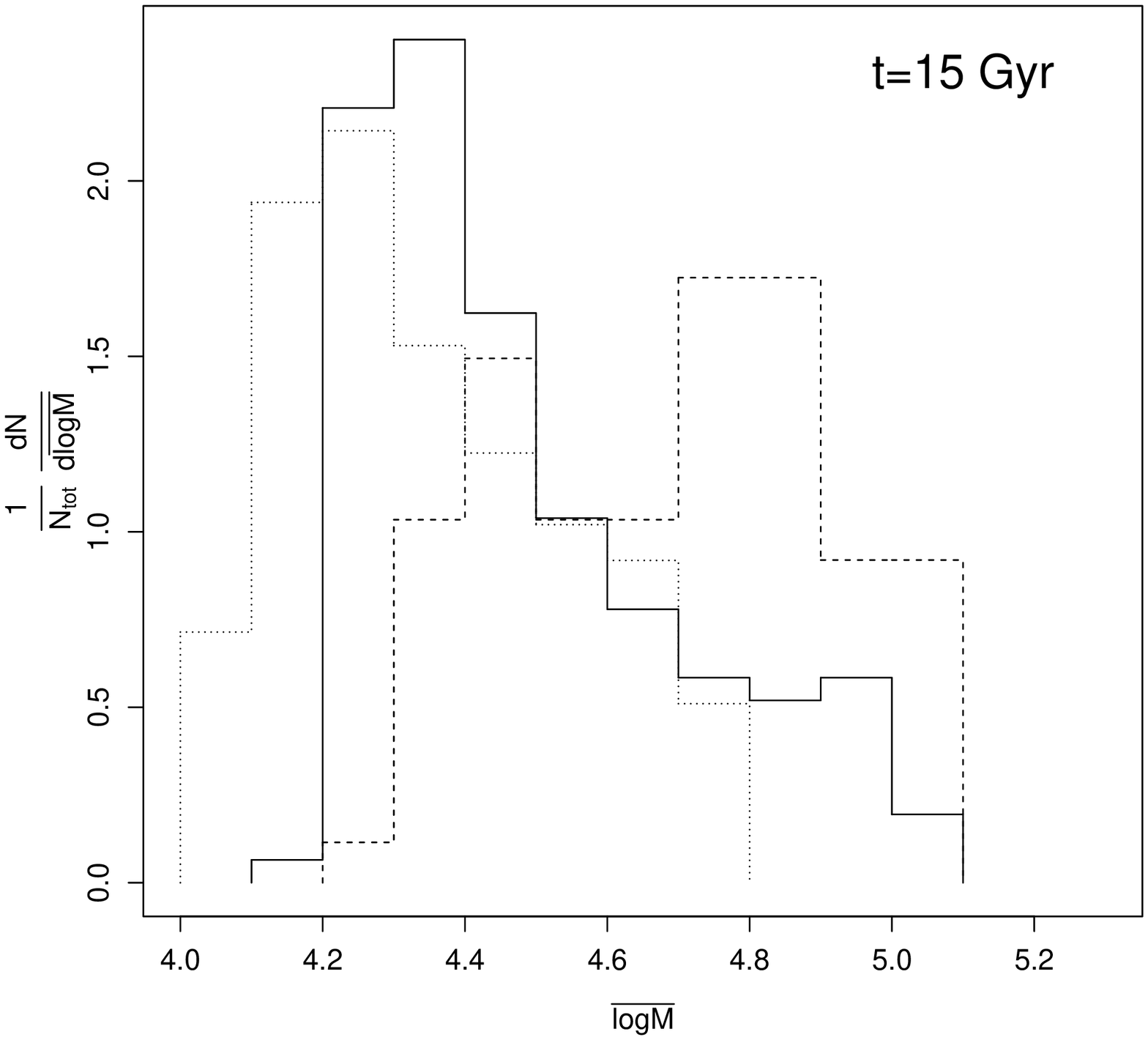,height=6cm,width=6cm,angle=0}}
\caption{ Evolution of the distribution of $\lm$ of globular cluster
systems located in host galaxies with values of
$R_e$ and $M_e$ plotted in Fig. 8. In each panel, the solid line shows
the distribution for globular cluster systems in host galaxies with $\log
M_e>10.5$, the dashed line that for  host galaxies
with $9.5<\log M_e<10.5$ and the dotted line that for host galaxies with $\log
M_e<9.5$. }
\end{figure*}
As shown in the observational analysis by
Harris (2000) (see also Whitmore 1997), clusters in
dwarf galaxies tend to have a mean 
mass lower than clusters in giant galaxies and a larger galaxy-to-galaxy
dispersion (in dwarf galaxies, for $M/L_V=2$, $\lmf=4.99$ with a
galaxy-to-galaxy dispersion equal to 0.24).
\begin{figure}
\centerline{\psfig{file=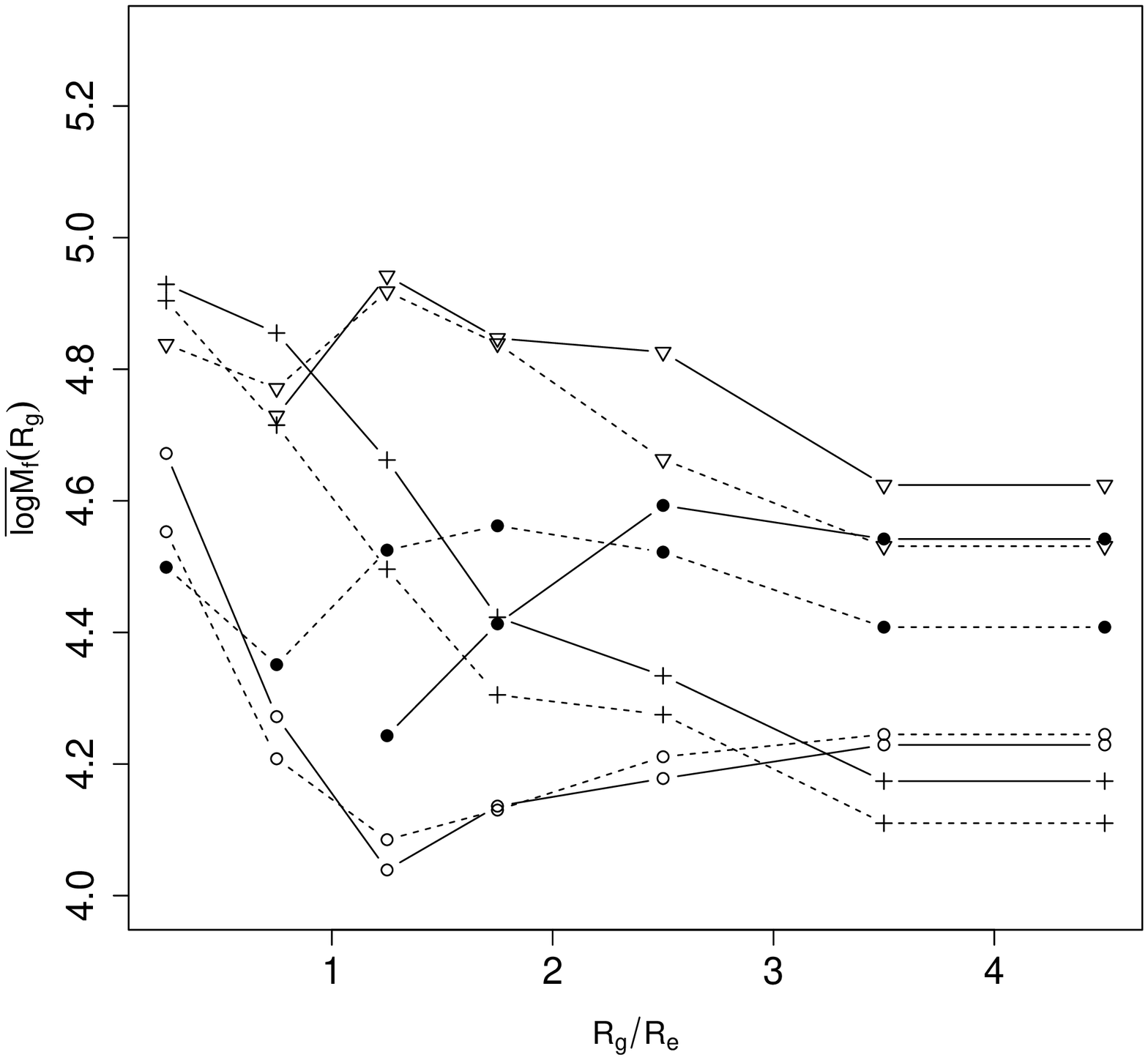,height=8cm,width=8cm,angle=0}}
\centerline{\psfig{file=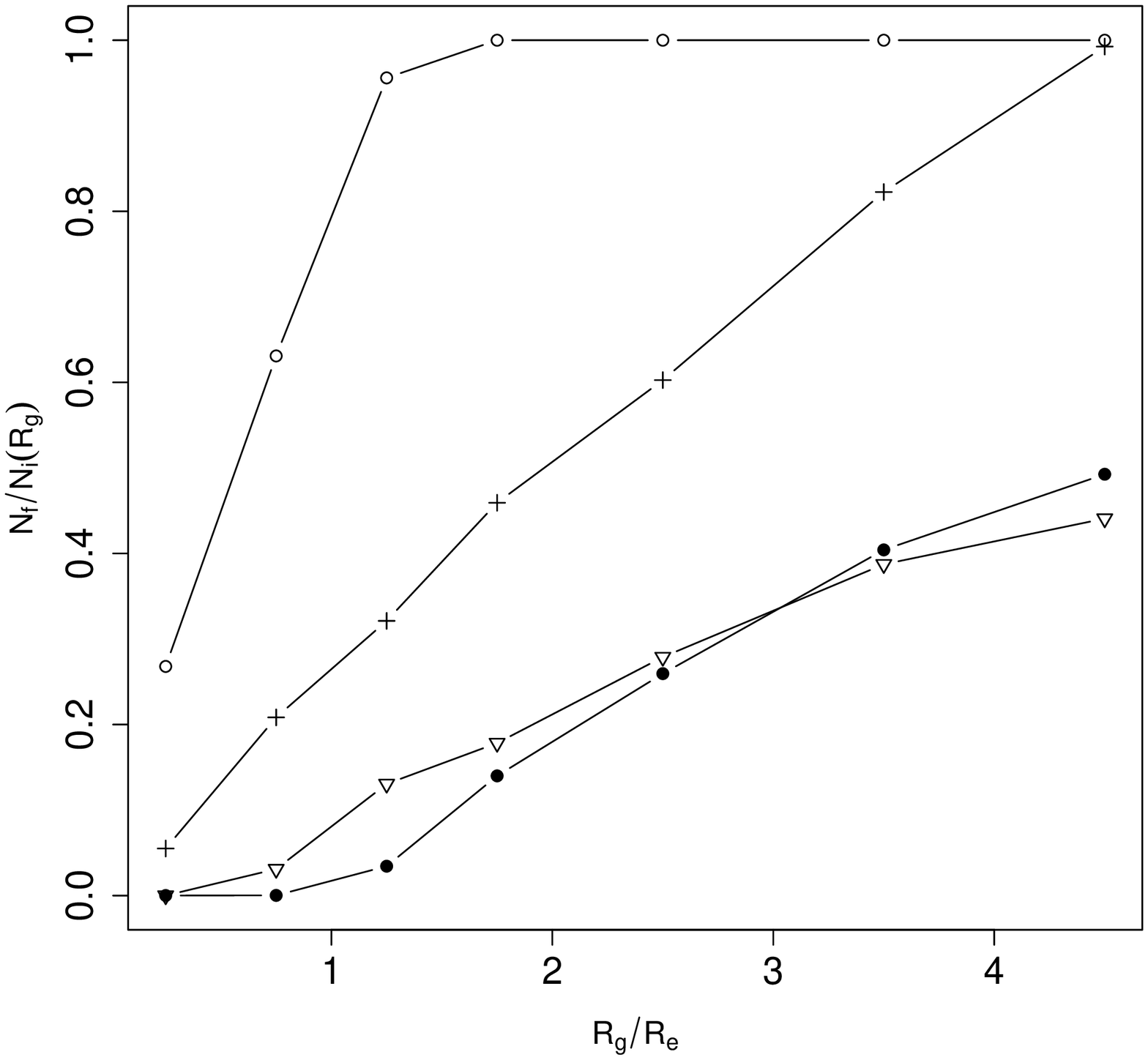,height=8cm,width=8cm,angle=0}}
\caption{$\lmf$ (upper panel) and $\nfni$ (lower panel) versus the
galactocentric distance (normalized by the  
effective radius of the host galaxy)  for globular cluster systems in host
galaxies with effective masses and radii plotted in Fig. 8 as filled
dots (open dots
are for $(\log M_e,\log R_e)=(12,1.4)$, crosses for $(\log M_e,\log
R_e)=(11,0.7)$, triangles for $(\log M_e,\log R_e)=(10,0.1)$ and filled
dots for $(\log M_e,\log R_e)=(9,-0.2)$).  
The results are from simulations with an initial power-law GCMF with
$\alpha=1.8$. The dashed lines in the upper panel show $\lmf$ versus
the projected galactocentric distance (normalized by the effective
radius of the host galaxy).} 
\end{figure}
\begin{figure}
\centerline{\psfig{file=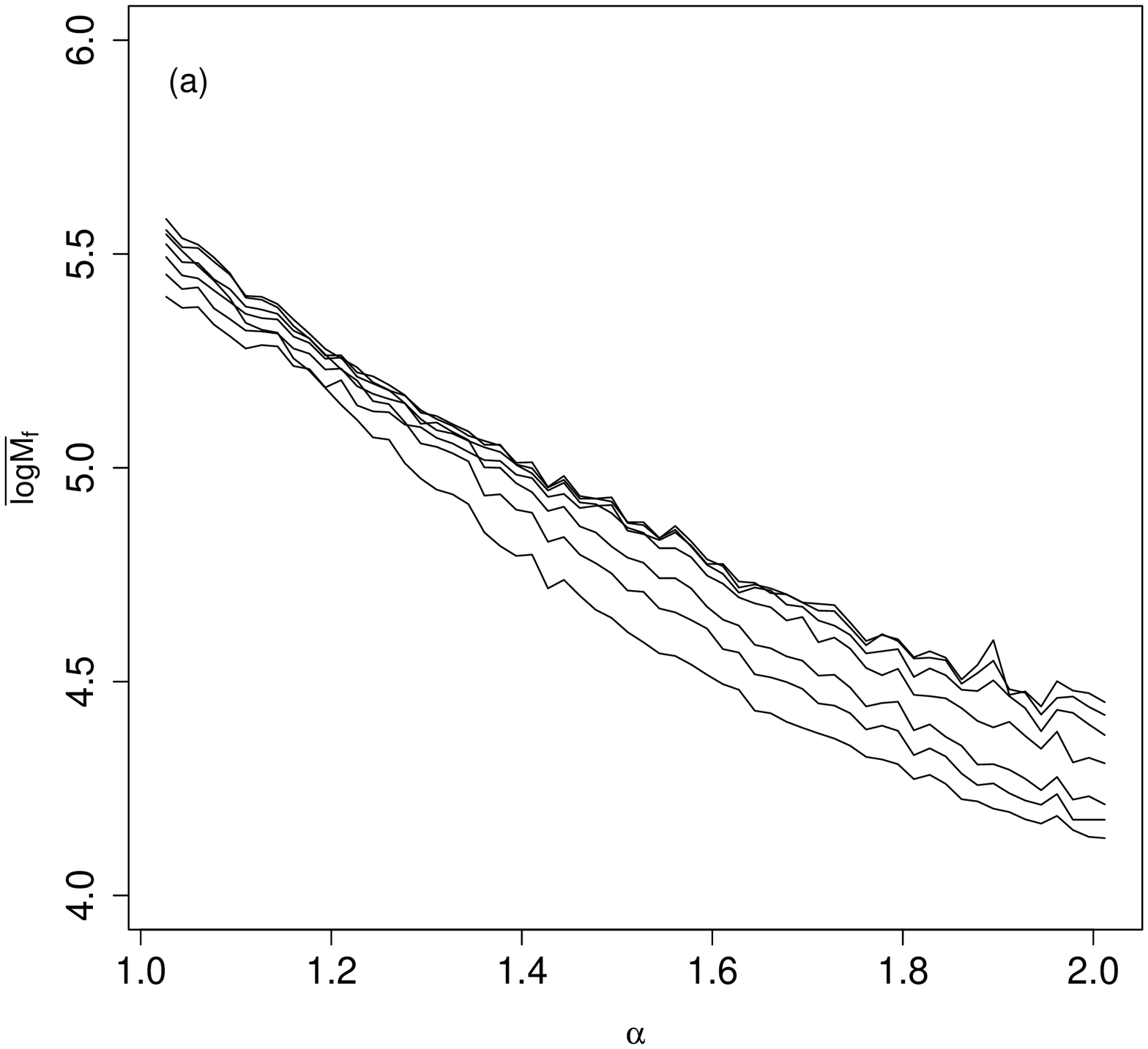,height=6cm,width=6cm,angle=0}}
\centerline{\psfig{file=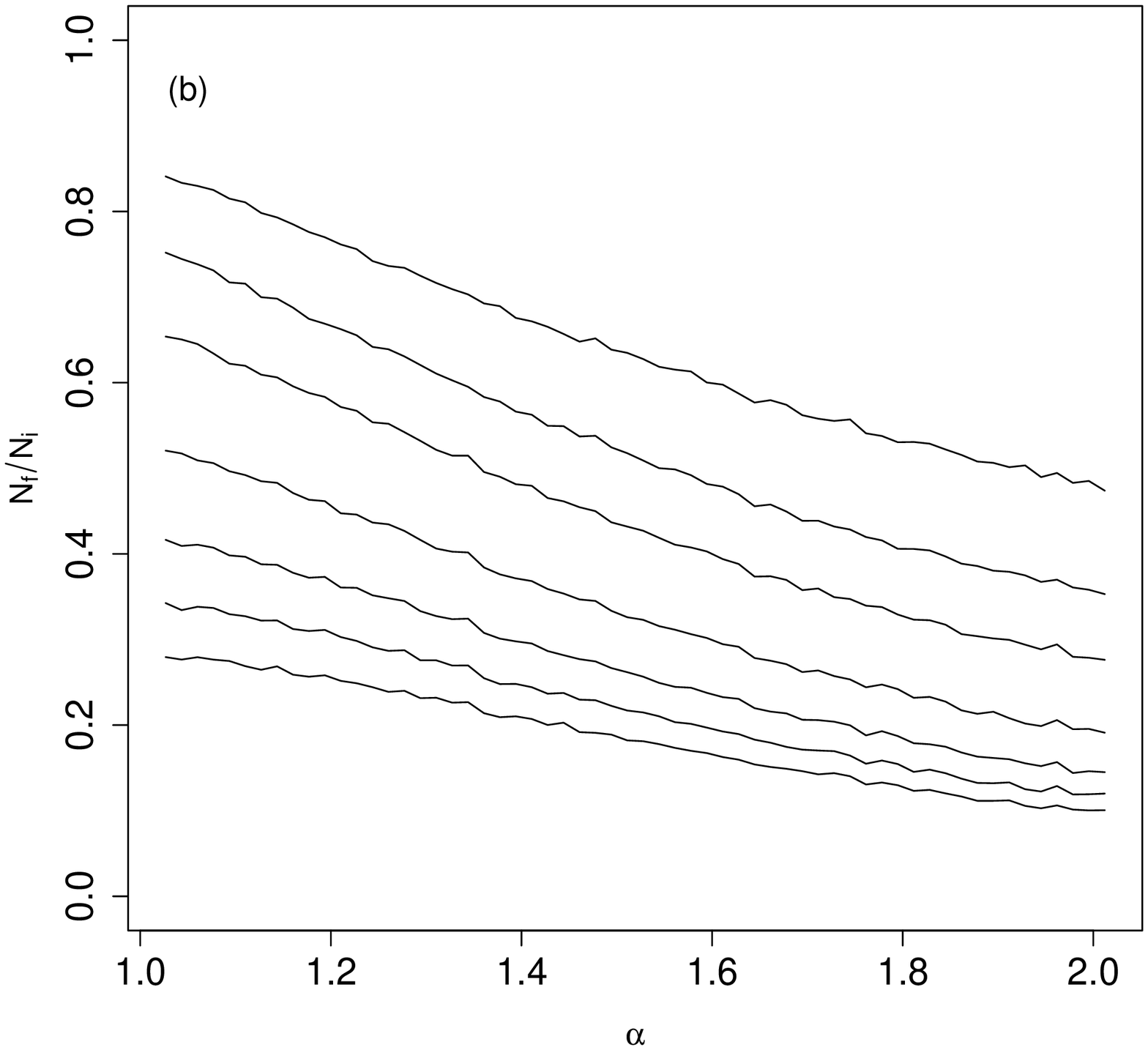,height=6cm,width=6cm,angle=0}}
\centerline{\psfig{file=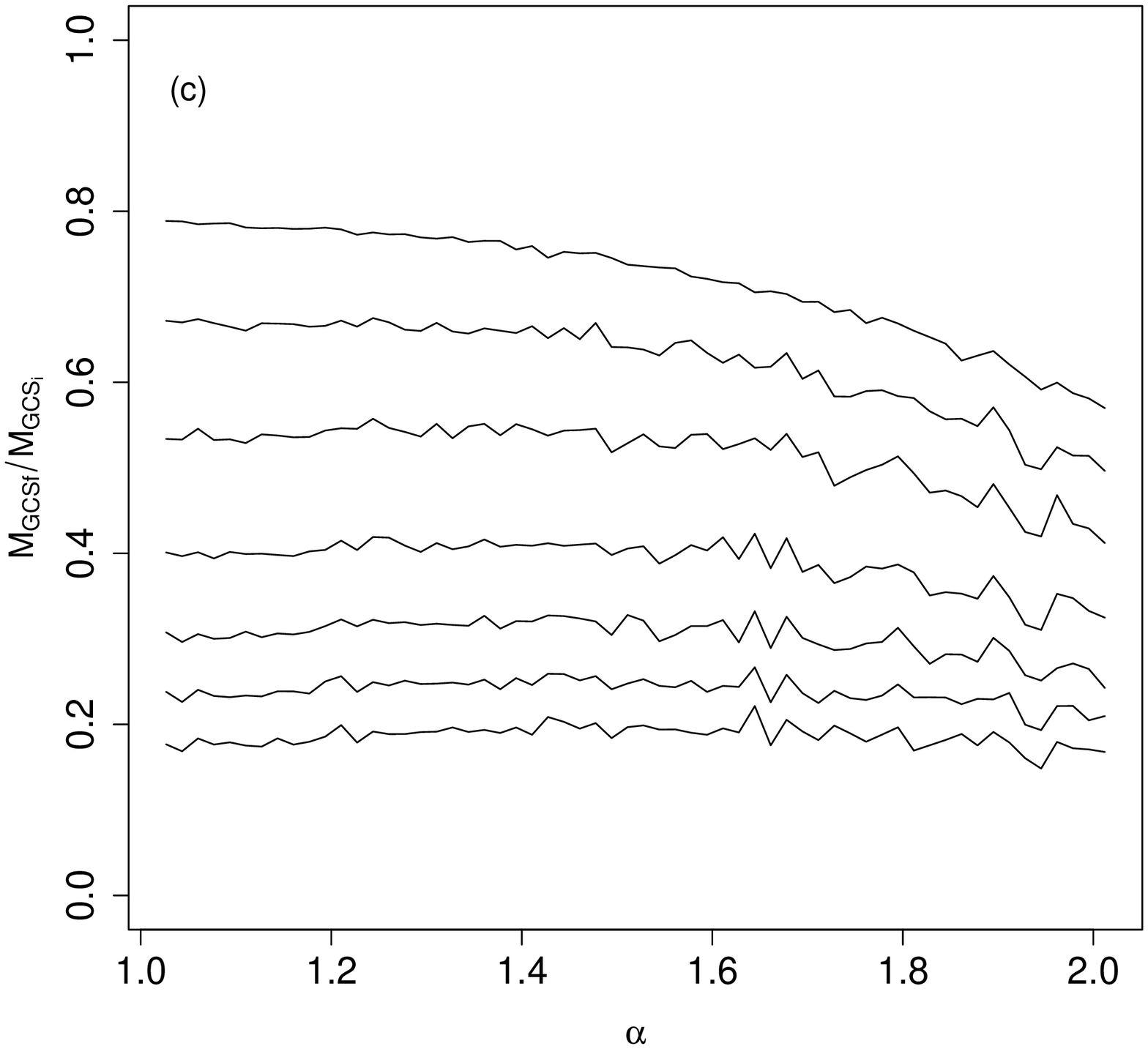,height=6cm,width=6cm,angle=0}}
\caption{(a) $\lmf$, (b) $\nfni$, (c) $\mfmi$ versus the index,
$\alpha$, of the initial power-law GCMF for the seven host galaxies
considered in section 4.3. In panels (b) and (c), the seven curves
shown 
correspond, from the upper to the lower one, to 
host galaxies with decreasing values of $M_e$; in panel (a) the curves
shown correspond, from the upper to the lower according to their
position on the right side of
the panel, to host galaxies with increasing values of  $M_e$.}
\end{figure}
It is clear from Fig.9 that, adopting a power-law initial GCMF, 
the final values of $\lmf$, the galaxy-to-galaxy dispersion of
$\lmf$ for different classes of galaxies, and the difference between
the distribution of $\lmf$ for dwarf and giant ellipticals are
not consistent with the current observational results.
The large scatter in the distribution of $\lmf$ for different
classes of galaxies, shown in Fig.9, is a consequence of the strong
dependence of the evolution of the GCMF on the structure of the host
galaxy when a power-law initial GCMF is adopted.

\subsection{Radial dependence of the GCMF}
For four fiducial systems with values of $\lre$ and $\lme$ shown as
filled dots in Fig. 8 we have studied in larger detail the radial
variation of $\lmf$ and of $\nfni$.

Fig. 10 shows $\lmf$ and $\nfni$ vs $R_g/R_e$. For all the four host
galaxies considered, the fraction of disrupted
clusters is large also beyond
$R_e$; only for the most massive galaxy there is no disruption beyond
$R_g=2R_e$ while for all the others disruption is significant at any
galactocentric distance. As shown in the upper panel of Fig.10, $\lmf$ varies
significantly with the galactocentric distance: the variation of
$\lmf$ with $R_g$ depends on the structure of the host galaxy and it
is not necessarily monotonic. For the two most massive host galaxies
considered, the mean mass of the innermost clusters and
that of the outermost clusters differ by approximately one order of
magnitude or slightly less.

Since in external galaxies only projected distances can be determined,
to ease the comparison with observations, the upper panel of Fig.10
shows also the variation of $\lmf$ with the projected
galactocentric distance. It is clear from Fig.10 that the radial
gradient of $\lmf$ is too large to be consistent with that reported by
observational studies.

It is interesting to note the contrast between the results shown in Fig.
10 and those 
obtained in Vesperini (2000) for GCS with a log-normal initial GCMF
(see Figs 12a and 12b in Vesperini 2000): with a log-normal initial
GCMF, although a 
 significant fraction of clusters is disrupted, the radial
gradient of $\lmf$ is, in most cases, weak and consistent
with observations. 

\subsection{Dependence of the results on the slope of the initial GCMF}
In the previous two subsections we have shown that the final
properties of GCS with a power-law initial
GCMF with a value of $\alpha$ similar to that observed in several young
cluster systems are not
consistent with the observational properties of old cluster systems. 
In particular the values of $\lmf$ for high-mass host galaxies  have
been found to be smaller than those observed and the galaxy-to-galaxy
dispersion much larger than that observed.

Although values of $\alpha$ too different
from that adopted above have not received any support  either by
theoretical studies of clusters formation or by observational analyses of
young cluster systems, it is interesting to study the dependence of
our results on 
$\alpha$ and 
explore the evolution of GCS with initial values of $\alpha$ outside
the range considered  in the previous sections. 

We focus our
attention on high-mass galaxies with $\lme>10.5$ and consider the
following values of ($\lme,~\lre$):(12,1.4), (11.75,1.18), (11.5,1),
(11.25,0.8), (11.0,0.64), (10.75,0.52), (10.5,0.4); 
these values span the entire strip of the $\log M_e-\log R_e$ plane covered by
real galaxies with $\log M_e>10.5$.
Figs 11a-c show $\lmf$, $\nfni$ and $\mfmi$ versus $\alpha$. As
expected, the smaller $\alpha$, the larger the fraction of
surviving clusters (but note that the disruption is never negligible)
and the larger the final values of $\lmf$. Values of $\lmf$ consistent
with those observed can result only from initial GCMF with values of
$\alpha$ ($\alpha\simeq 1.2-1.3$) significantly flatter than those observed in
young cluster systems.
\section{Summary and conclusions}
We have investigated the evolution of the main properties of GCS in
elliptical galaxies starting with an initial power-law GCMF 
similar to that of young cluster systems observed in merging galaxies.

We have considered a large set of different
host galaxies and we have shown the dependence of the final parameters
of the GCMF, of the fraction of surviving clusters and of the ratio of
the final total mass in clusters to the total initial mass of clusters
on the effective radius and effective mass of the host galaxy (see
Figs 2, 3, and 4). Three different values (close to those determined
observationally from the GCLF of young cluster systems) of the index of the
initial power-law GCMF have been explored ($\alpha=1.5,~1.8,~2.0$).

While evolutionary
processes can easily turn a GCMF with an 
initial power-law shape into a GCMF with a log-normal
shape like that observed in old cluster systems, our investigation
reveals several inconsistencies between the observed properties
of old GCS in elliptical galaxies and the theoretical
results obtained adopting a power-law initial GCMF. In particular,
studying the evolution of GCS in a sample of giant, normal and dwarf
elliptical galaxies with effective radii and effective masses equal to
those determined observationally, we have shown that:
\begin{itemize}
\item the theoretical values of $\lmf$ are in
general  smaller than those observed;
\item the range spanned by $\lmf$ is  significantly larger than that
reported by observational studies; in particular, for galaxies with
$\lme>10.5$, we have shown that starting with a power-law initial GCMF
with $\alpha=1.8$, $\lmf$ ranges from 4.2 to 5.0 whereas  observations
show an approximately universal 
value of $\lmf$ ($\lmf \simeq 5.16$ for $M/L_V=2$) with a very small 
galaxy-to-galaxy dispersion ($\sim 0.06$);
\item $\lmf$ varies significantly with the effective mass of the host
galaxy; specifically $\lmf$  tends to increase with decreasing values
of $\lme$ for $\lme>10$, it reaches a peak and then decreases again for
dwarf galaxies; this does not agree with observations which show that
for giant galaxies, $\lmf$ is approximately constant and larger than
$\lmf$ of clusters in dwarf galaxies. 
\item adopting a power-law initial GCMF,
a significant dependence of $\lmf$ on the galactocentric distance is
produced by the effects of evolutionary processes; this is in contrast 
with several observational studies which fail to
find a significant radial gradient of $\lmf$ within individual
galaxies.
\item Starting with an initial power-law GCMF, values of $\lmf$
consistent with those observed in massive  galaxies can be obtained
only with values of $\alpha$ smaller ($\alpha \simeq 1.2-1.3$) than
those observed in young cluster systems.
\end{itemize}
In Vesperini (2000) it has been shown that the final properties of GCS obtained
starting with a log-normal initial GCMF similar to that observed in
the external regions of some elliptical galaxies (where evolutionary
processes are unlikely to have significantly altered the initial
conditions of clusters)  are in general good  agreement
with the observed properties of old cluster systems. 
In this paper we have shown that,
starting with a power-law initial GCMF, the final GCS properties present
several discrepancies from those observed in old GCS; this would seem to
rule out the possibility that old cluster systems were formed with a power-law
initial GCMF similar to that observed in young cluster systems of
merging galaxies and to strongly favour a log-normal initial GCMF.

One can 
not exclude the existence of differences in the process of globular
cluster formation at the current epoch and at the time of formation of
currently old clusters leading to different initial GCMF.
While individual young clusters could indeed be young globular
clusters and  those which survive could evolve into systems
similar to old globular clusters, 
the future global properties of these young cluster systems could
differ from
those of currently old cluster systems. 
\section*{ACKNOWLEDGMENTS}
I wish to thank an anonymous referee for useful comments on the paper.
Support from a Five College Astronomy Department fellowship is acknowledged.
\section*{References}
Aguilar L. , Hut P.  Ostriker J.P. 1988, ApJ, 335, 720\\
Baumgardt H., 1998, A\&A, 330, 480\\
Binney J., Tremaine S., 1987, Galactic Dynamics, Princeton University
Press, Princeton, New Jersey\\ 
Burstein D. , Bender R., Faber S., Nolthenius R., 1997, AJ, 114, 1365\\
Caputo F., Castellani V., 1984, MNRAS, 207,185\\
Capuzzo Dolcetta R., Tesseri A., 1997, MNRAS, 292, 808\\
Carlson M.C., et al. 1998, AJ, 115, 1778\\
Carlson M.C., et al. 1999, AJ, 117, 1700\\
Chernoff D.F., Kochanek C.S., Shapiro S.L., 1986, ApJ, 309, 183\\
Chernoff D.F., Shapiro S.L., 1987, ApJ, 322, 113 \\
Chernoff D.F., Weinberg M.D., 1990, ApJ, 351, 121\\
Elmegreen B.G., Efremov Y.N., 1997, ApJ, 480, 235\\
Fall S.M., Rees M.J., 1977, MNRAS, 181, 37P\\
Fall S.M., Malkan M.A., 1978, MNRAS, 185, 899\\
Fritze-von Alvensleben U., 1998, AA, 336, 83\\
Fritze-von Alvensleben U., 1999, AA, 342, L25\\
Gnedin O.Y., Ostriker J.P., 1997, ApJ, 474, 223\\
Gnedin O.Y., Hernquist L., Ostriker J.P., 1999, ApJ, 514, 109\\
Harris W.E., 2000, in Lectures for 1998 Saas-Fee Advanced School on Star
Clusters, in press\\
Harris W.E., Pudritz R.E., 1994, ApJ, 429, 177\\
Johnson K.E., Vacca W.D., Leitherer C., Conti P.S., Lipscy S.J., 1999,
AJ, 117, 1708\\
Miller B.W.,  Whitmore B.C.,Schweizer F.,Fall S.M., 1997, AJ, 114, 2381 \\
Murali C., Weinberg M.D., 1997a, MNRAS, 288, 767\\
Murali C., Weinberg M.D., 1997b, MNRAS, 291, 717\\
Ho L.C., Filippenko A.V., 1996a, ApJ, 466, L83\\
Ho L.C., Filippenko A.V., 1996b, ApJ, 472, 600\\
Okazaki, T., Tosa, M., 1995, MNRAS, 274, 48\\
Ostriker J.P., Gnedin O.Y., 1997, 487, 667\\
Schweizer F., Miller B.W., Whitmore B.C., Fall S.M., 1996, AJ, 112, 1839\\
van den Bergh S., 1995b, AJ, 110, 2700\\
Vesperini E., 1994, Ph.D. Thesis, Scuola Normale Superiore, Pisa, Italy\\
Vesperini E., 1997, MNRAS, 287, 915\\
Vesperini E., 1998, MNRAS, 299, 1019\\
Vesperini E., 2000, MNRAS, in press;\\
Vesperini E., Heggie D.C., 1997, MNRAS, 289, 898\\
Weinberg M.D., 1994a, AJ, 108,1398\\
Weinberg M.D., 1994b, AJ, 108, 1403\\
Weinberg M.D., 1994c, AJ, 108, 1414\\
Whitmore B.C., 1999, in Galaxy interactions at low and high redshift,
IAU Symp. 186, Barnes J.E. \& Sanders D.B. editors, Kluwer Academic
Publishers, p.251\\
Whitmore B.C., 1997, in The Extragalactic distance scale, M.Livio,
M.Donahue, N.Panagia editors, Cambridge University Press\\ 
Whitmore B.C., Zhang Q., Leitherer C., Fall S.M., Schweizer F., Miller
B.W., 1999, AJ, 118, 1551\\
Zepf S.E., Ashman K.M., English J., Freeman K.C., Sharples R.M., 1999,
AJ, 118, 752 \\ 
Zhang Q., Fall S.M., 1999, ApJ, 527, L81\\

\end{document}